\documentclass[12pt]{amsart}
\usepackage{epsfig}

\begin{document}
\setcounter{page}{1}

\newcommand{\erf}{\mathop{\rm erf}\nolimits}

\title[Analytic error function and numeric inverse
\dots]{Analytic error function and numeric inverse obtained by geometric means}
\author[Dmitri Martila]{Dmitri Martila}
\author[Stefan Groote]{Stefan Groote}
\address{Institute of Physics, University of Tartu, Estonia}
\thanks{eestidima@gmail.com, stefan.groote@ut.ee}

\begin{abstract}
Using geometric considerations, we provide a clear derivation of the integral
representation for the error function, known as the Craig formula. We calculate
the corresponding power series expansion and prove the convergence. The same
geometric means finally help to systematically derive handy formulas that
approximate the inverse error function. Our approach can be used for
applications in e.g.\ high-speed Monte Carlo simulations where this function
is used extensively.\\[12pt]
MSC Class: 62E15, 62E17, 60E15, 26D15.
\end{abstract}

\maketitle

\section{Introduction}
High-speed Monte Carlo simulations are used for a large spectrum of
applications from mathematics to economy. As input for such simulations, the
probability distribution are usually generated by pseudo-random number
sambling, a method going back to a work of John von Neumann from
1951~\cite{vonNeumann:1951}. In the era of ``big data'' such methods have to
be fast and reliable, a sign of such neccessity being the release of the very
first randomness Quside processing unit in 2023~\cite{111}. Still, these
samblings need to be cross-checked by exact methods, and for these the
knowledge of analytical functions to describe the stochastic processes, among
those the error function, are of tremendous importance.

By definition, a function is called analytic if it locally given by a
converging Taylor series expansion. Even if a function itself turns out not
to be analytic, its inverse can be analytic. The error function can be given
analytically, one of these analytic expressions is the integral representation
given by Craig in 1991~\cite{Craig:1991}. Craig mentioned this representation
only in passing and did not give a derivation of it. In the following, there
have been a couple of derivations of this formula~\cite{Lever:1998,%
Tellambura:1999,Stewart:2017}. In Sec.~2 we add a further one which is based
on the same geometric considerations as employed in Ref.~\cite{Martila:2022}.
In Sec.~3 we give the series expansion for Craig's integral representation and
show the fast convergence of this series.

For the inverse error function, handbooks for special functions (cf.\ e.g.\
Ref.~\cite{book}) do not unveil such an analytic property. Instead, this
function have to be approximated. Known approximations are dating back to the
late 1960s and early 1970s~\cite{Strecok:1968,Blair:1976}) and reach up to
semi-analytical approximations by asymptotic expansion (cf., e.g.,
Refs.~\cite{Bergsma:2006,Dominici:2007,Dominici:2008,Winitzki:2008,Giles:2011,%
Soranzo:2012}. Using the same geometric considerations, in Sec.~4 we develop a
couple of handy approximations which can easily be implemented in different
computer languages, indicating the deviations from an exact treatment. In
Sec.~5 we discuss our result and test the CPU time. Sec.~6 contains our
conclusions.

\section{Derivation of Craig's integral representation}
Ref.~\cite{Martila:2022} provides an approximation for the integral over the
Gaussian standard normal distribution obtained by geometric considerations
that is related to the cumulative distribution function via
$P(t)=\Phi(t)-\Phi(-t)$, where $\Phi(t)$ is the Laplace function. The same
considerations apply to the error function $\erf(t)$ which is related to
$P(t)$ via
\begin{equation}\label{erf}
\erf(t)=\frac1{\sqrt\pi}\int_{-t}^te^{-x^2}dx
  =\frac1{\sqrt{2\pi}}\int_{-\sqrt2t}^{\sqrt2t}e^{-x^2/2}dx=P(\sqrt2t).
\end{equation}
Translating the results of Ref.~\cite{Martila:2022} to the error function, one
obtains the approximation of order $p$ to be
\begin{equation}\label{Eqf2}
\erf_p(t)^2=1-\frac1N\,\sum_{n=1}^Ne^{-k_{p,n}^2t^2},
\end{equation}
where the $N=2^p$ values $k_{p,n}$ ($n=1,2,\ldots,N$) are found in the
intervals between $1/\cos(\pi(n-1)/(4N))$ and $1/\cos(\pi n/(4N))$. The way
of selecting those values is extensively described in Ref.~\cite{Martila:2022}
where it is shown that
\begin{equation}\label{rteq}
\Big|\erf(t)-\sqrt{1-e^{-k_{0,1}^2t^2}}\Big|<0.0033
\end{equation}
for $k_{0,1}=1.116$, and with $14\approx 0.0033/0.00024$ times larger
precision
\begin{equation}\label{rteq2}
\Big|\erf(t)-\sqrt{1-\frac12(e^{-k_{1,1}^2t^2}+e^{-k_{1,2}^2t^2})}\Big|<0.00024,
\end{equation}
for $k_{1,1}=1.01$, $k_{1,2}=1.23345$. For the parameters taking the values
$k_{p,n}=1/\cos(\pi n/(4N))$ of the upper limits of those intervals, it can be
shown that the deviation is given by
\begin{equation}\label{Eqf299}
|\erf(t)-\erf_p(t)|<\frac{\exp(-t^2)}{2N}\,\sqrt{1-\exp(-t^2)}\,.
\end{equation}
Given the values $k_{p,n}=1/\cos\phi(n)$ with $\phi(n)=\pi n/(4N)$, in the
limit $N\to\infty$ the sum over $n$ in Eq.~(\ref{Eqf2}) can be replaced by an
integral with measure $dn=(4N/\pi)d\phi(n)$ to obtain
\begin{equation}\label{A6}
\erf(t)^2=1-\frac4\pi\int_0^{\pi/4}\exp\left(\frac{-t^2}{\cos^2\phi}\right)
  \,d\phi.
\end{equation}

\section{Power series expansion}
The integral in Eq.~(\ref{A6}) can be expanded into a power series in $t^2$,
\begin{equation}\label{A7}
\erf(t)^2=1-\frac4\pi\sum_{n=0}^\infty c_n\frac{(-1)^n}{n!}(t^2)^n
\end{equation}
with
\begin{eqnarray}\label{cn}
c_n&=&\int_0^{\pi/4}\frac{d\phi}{\cos^{2n}\phi}\ =\ \int_0^{\pi/4}
  (1+\tan^2\phi)^nd\phi\ =\ \int_0^1(1+y^2)^{n-1}dy\nonumber\\
  &=&\sum_{k=0}^{n-1}\begin{pmatrix}n-1\\k\\\end{pmatrix}\int_0^1y^{2k}dy
  \ =\ \sum_{k=0}^{n-1}\frac{1}{2k+1}\begin{pmatrix}n-1\\ k\\\end{pmatrix},
\end{eqnarray}
where $y=\tan\phi$. The coefficients $c_n$ can be expressed by the
hypergeometric function, $c_n={}_2F_1(1/2,1-n;3/2;-1)$, also known as Barnes'
extended hypergeometric function. On the other hand, we can derive a
constraint for the explicit finite series expression for $c_n$ that renders
the series in Eq.~(\ref{A7}) to be convergent for all values of $t$. In order
to be self-contained, intermediate steps to derive this constraint and to show
the convergence are shown in the following. Necessary is Pascal's rule
\begin{eqnarray}
\lefteqn{\begin{pmatrix}n\\k\end{pmatrix}+\begin{pmatrix}n\\k-1\end{pmatrix}
  \ =\ \frac{n!}{k!(n-k)!}+\frac{n!}{(k-1)!(n-k+1)!}}\nonumber\\
  &=&\frac{n!(n-k+1+k)}{k!(n-k+1)!}\ =\ \frac{(n+1)!}{k!(n+1-k)!}
  \ =\ \begin{pmatrix}n+1\\k\end{pmatrix}
\end{eqnarray}
and the sum over the rows of Pascal's triangle,
\begin{equation}\label{row}
\sum_{k=0}^n\begin{pmatrix}n\\k\end{pmatrix}=2^n
\end{equation}
which can be shown by mathematical induction. The base case $n=0$ is obvious,
as $(\begin{smallmatrix}0\\0\end{smallmatrix})=1=2^0$. For the induction step
from $n$ to $n+1$ we write the first and last elements
$(\begin{smallmatrix}n+1\\0\end{smallmatrix})=1$ and
$(\begin{smallmatrix}n+1\\n+1\end{smallmatrix})=1$ separately and use
Pascal's rule to obtain
\begin{eqnarray}
\lefteqn{\sum_{k=0}^{n+1}\begin{pmatrix}n+1\\k\end{pmatrix}
  \ =\ 1+\sum_{k=1}^n\begin{pmatrix}n+1\\k\end{pmatrix}
  +1\ =}\nonumber\\
  &=&1+\sum_{k=1}^n\begin{pmatrix}n\\k\end{pmatrix}
  +\sum_{k=1}^n\begin{pmatrix}n\\k-1\end{pmatrix}+1
  \ =\ 2\sum_{k=0}^n\begin{pmatrix}n\\k\end{pmatrix}\ =\ 2^{n+1}.
\end{eqnarray}
This proves Eq.~(\ref{row}). Returning to Eq.~(\ref{cn}), one has
$0\le k\le n-1$ and, therefore,
\begin{equation}
\frac1{2n-1}\le\frac1{2k+1}\le 1.
\end{equation}
For the result in Eq.~(\ref{cn}) this means that
\begin{equation}
\frac1{2n-1}\sum_{k=0}^{n-1}\begin{pmatrix}n-1\\k\end{pmatrix}\le c_n\le
\sum_{k=0}^{n-1}\begin{pmatrix}n-1\\k\end{pmatrix}=2^{n-1},
\end{equation}
i.e., the existence of a real number $c_n^*$ between $1/(2n-1)$ and $1$ such
that $c_n=c_n^*2^{n-1}$. One has
\begin{equation}
\erf_p(t)^2=1-\frac4\pi\sum_{n=0}^Nc_n\frac{(-1)^n}{n!}(t^2)^n
  =1-\frac2\pi\sum_{n=0}^Nc_n^*\frac{(-2t^2)^n}{n!},
\end{equation}
and because of $0\le c_n^*\le 1$ there is again a real number $c_N^{**}$ in
the corresponding open interval so that
\begin{equation}
\frac2\pi\sum_{n=0}^Nc_n^*\frac{(-2t^2)^n}{n!}=c_N^{**}\frac2\pi\sum_{n=0}^N
\frac{(-2t^2)^n}{n!}<\frac2\pi\sum_{n=0}^N\frac{(-2t^2)^n}{n!}.
\end{equation}
As the latter is the power series expansion of $(2/\pi)e^{-2t^2}$ which is
convergent for all values of $t$, also the original series is convergent and,
therefore, $\erf_p(t)^2$ with the limiting value shown in Eq.~(\ref{A7}). A
more compact form of the power series expansion is given by
\begin{equation}
\erf(t)^2=\sum_{n=1}^\infty c_n\frac{(-1)^{n-1}}{n!}(t^2)^n,\qquad
c_n=\sum_{k=0}^{n-1}\frac{1}{2k+1}\begin{pmatrix}n-1\\ k\\\end{pmatrix}.
\end{equation}

\section{Approximations for the inverse error function}
Based on the geometric approach from Ref.~\cite{Martila:2022}, in the
following we describe how to find simple, handy formulas that, guided by
higher and higher orders of the approximation~(\ref{Eqf2}) for the error
function lead to more and more advanced approximation of the inverse error
function. Starting point is the degree $p=0$, i.e., the approximation in
Eq.~(\ref{rteq}). Inverting $E=\erf_0(t)=(1-e^{-k_{0,1}^2t^2})^{1/2}$ leads
to $t^2=-\ln(1-E^2)/k_{0,1}^2$, and using the parameter $k_{0,1}=1.116$ from
Eq.~(\ref{rteq}) gives
\begin{equation}
T_0=\sqrt{-\ln(1-E^2)}/k_{0,1}^2.
\end{equation}
For $0\le E \le 0.92$ the relative deviation $(T_{(0)}-t)/t$ from the exact
value $t$ is less than $1.11\%$, for $0\le E<1$ the deviation is less than
$10\%$. Therefore, for $E>0.92$ a more precise formula has to be used. As such
higher values for $E$ appear only in $8\%$ of the cases, this will not
essentially influence the CPU time.

\begin{figure}[ht]\begin{center}
\epsfig{figure=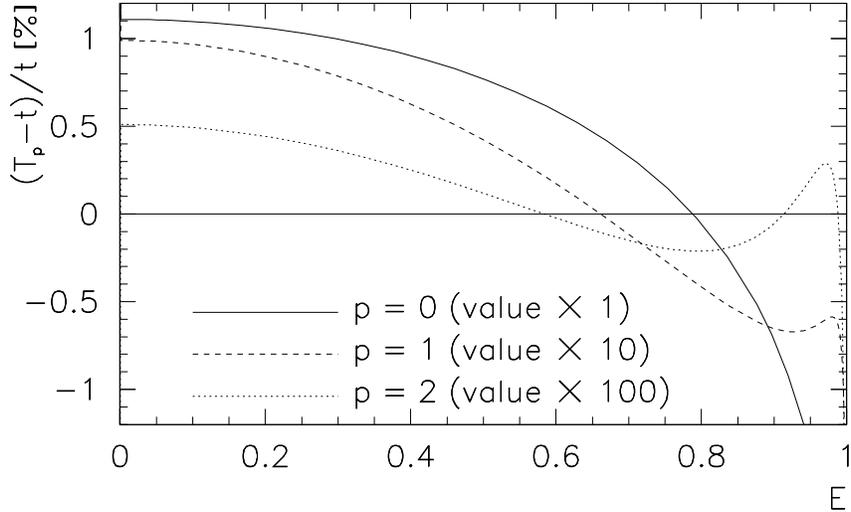, scale=0.7}
\caption{\label{rval012}Relative deviations for the statical approximations}
\end{center}\end{figure}

Continuing with $p=1$, we insert $T_0=\sqrt{-\ln(1-E^2)}/k_{0,1}^2$ into
Eq.~(\ref{Eqf2}) to obtain
\begin{equation}
\erf_1(T_0)=\sqrt{1-\frac12(e^{-k_{1,1}^2T_0^2}+e^{-k_{1,2}^2T_0^2})},
\end{equation}
where $k_{1,1}=1.01$ and $k_{1,2}=1.23345$ are the same as for
Eq.~(\ref{rteq2}). Taking the derivative of Eq.~(\ref{erf}) and approximating
this by the difference quotient, one obtains
\begin{equation}
\frac{\erf(t)-\erf(T_0)}{t-T_0}=\frac{\Delta\erf(t)}{\Delta t}\Big|_{t=T_0}
  \approx\frac{d\erf(t)}{dt}\Big|_{t=T_0}=\frac2{\sqrt\pi}e^{-T_0^2},
\end{equation}
leading to $t\approx T_1=T_0+\frac12\sqrt\pi e^{T_0^2}(E-\erf_1(T_0))$. In
this case, for in the larger interval $0\le E\le 0.995$ the relative deviation
$(T_1-t)/t$ is less than $0.1\%$. Using $\erf_2(t)$ instead of $\erf_1(t)$ and
inserting $T_1$ instead of $T_0$ one obtains $T_2$ with a relative deviation
of maximally $0.01\%$ for the same interval. The results are shown in
Fig.~\ref{rval012}.

\begin{figure}[ht]\begin{center}
\epsfig{figure=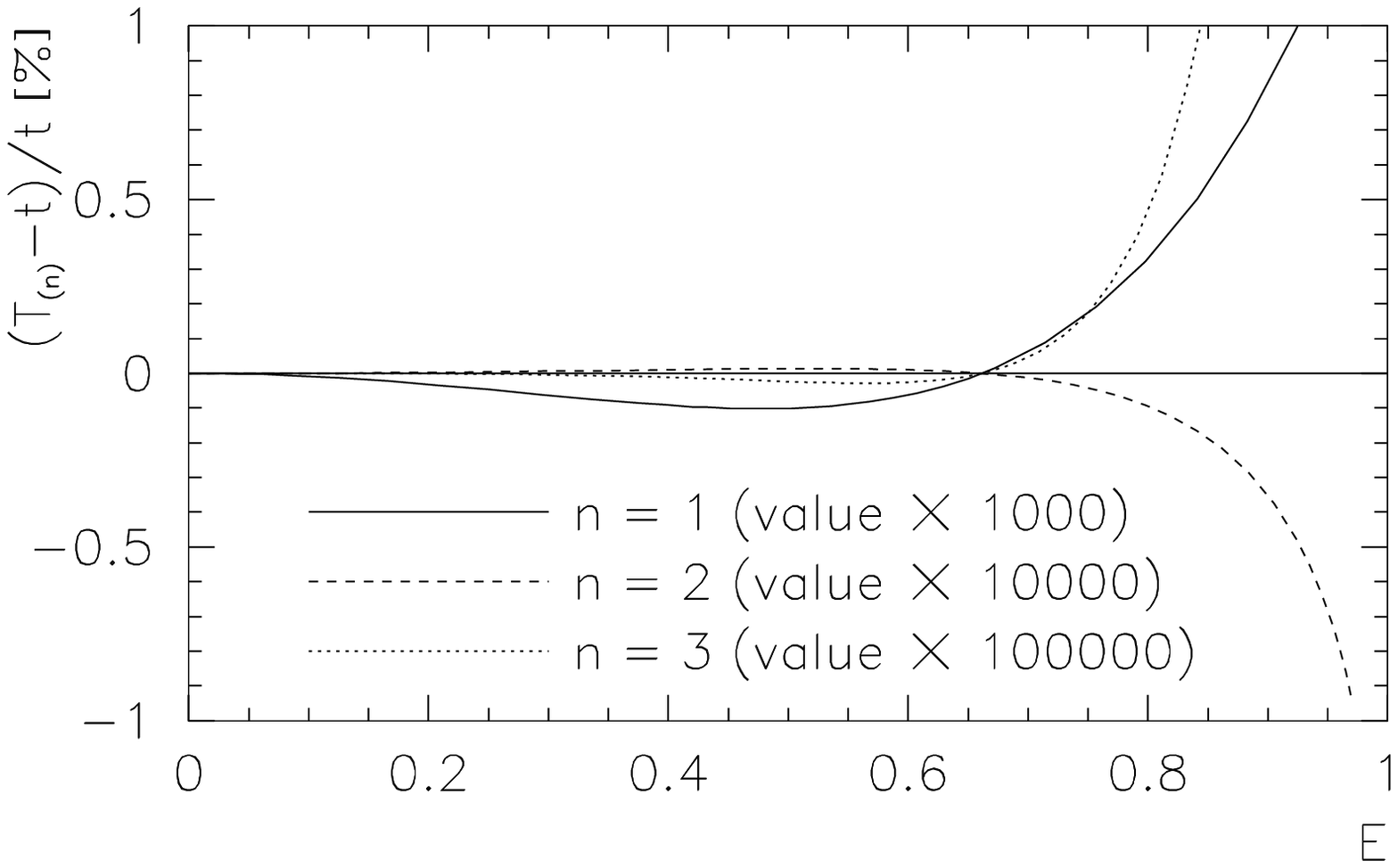, scale=0.7}
\vspace{7pt}
\epsfig{figure=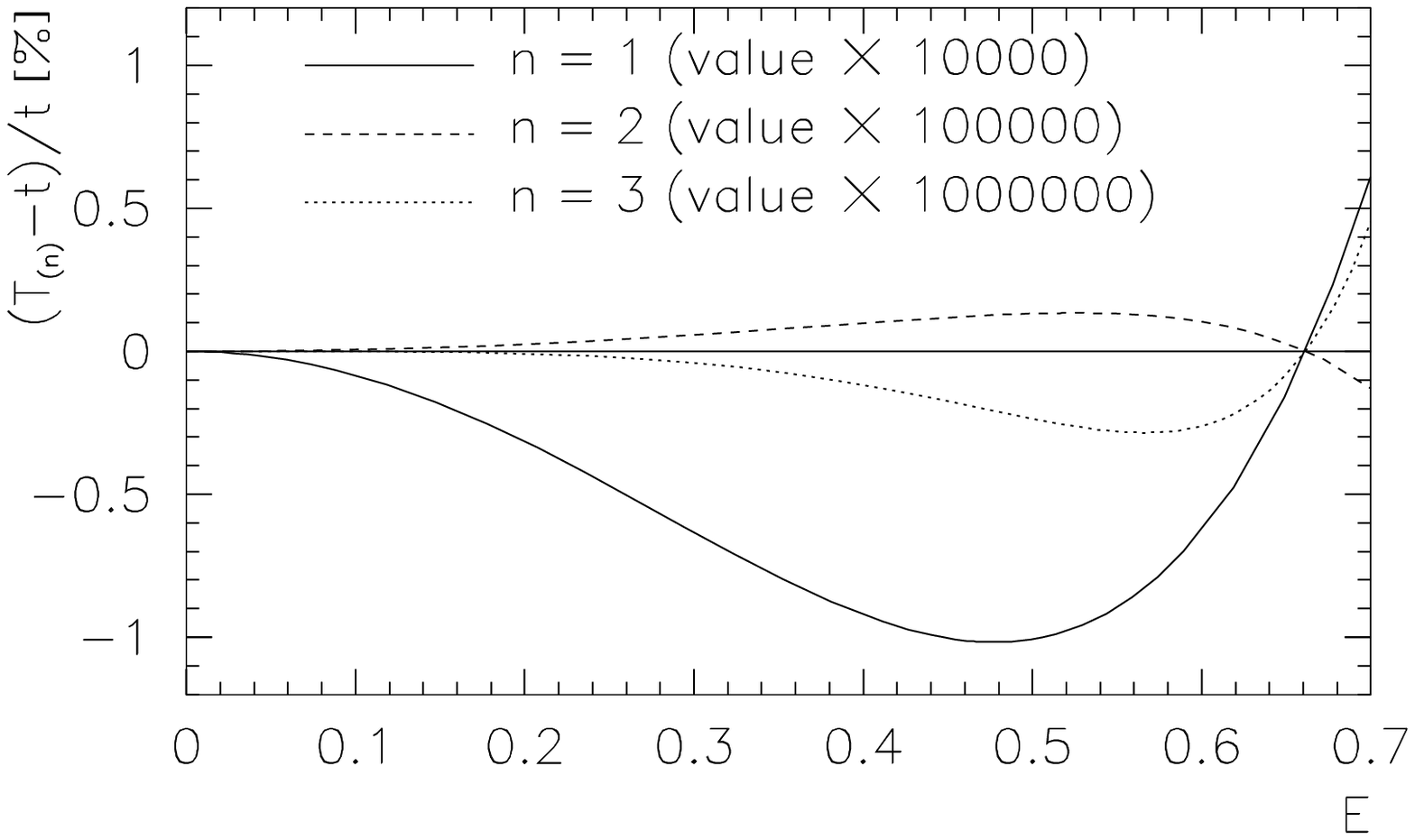, scale=0.7}
\caption{\label{rval456}Relative deviation for the dynamical approximations
  (the degree is chosen to be $p=1$)}
\end{center}\end{figure}

The method to can be optimised by a method similar to the shooting method in
boundary problems, giving dynamics to the calculation. Suppose that following
one of the previous methods, for a particular argument $E$ we have found an
approximation $t_0$ for the value of the inverse error function at this
argument. Using $t_1=1.01t_0$, one can adjust the improved result
\begin{equation}
t=t_0+A(E-\erf(t_0))
\end{equation}
by inserting $E=\erf(t)$ and and calculating $A$ for $t=t_1$. In general, this
procedure gives a vanishing deviation close to $E=0$. In this case and for
$t_0=T_1$, in the interval $0\le E\le 0.7$ the maximal deviation is slightly
larger than $10^{-6}=0.0001\%$ while up to $E=0.92$ the deviation is
restricted to $10^{-5}=0.001\%$. A more general ansatz
\begin{equation}
t=t_0+A(E-\erf(t_0))+B(E-\erf(t_0))^2
\end{equation}
can be adjusted by inserting $E=\erf(t)$ for $t=1.01t_0$ and $t=1.02t_0$, and
the system of equations
\begin{equation}
\Delta t=A\Delta E_1+B\Delta E_1^2,\qquad
2\Delta t=A\Delta E_2+B\Delta E_2^2
\end{equation}
with $\Delta t=0.01t_0$, $\Delta E_i=\erf(t_i)-\erf(t_0)$ can be solved for
$A$ and $B$ to obtain
\begin{equation}
A=-\frac{(2\Delta E_1^2-\Delta E_2^2)\Delta t}{\Delta E_1\Delta E_2
  (\Delta E_1-\Delta E_2)},\quad
B=\frac{(-2\Delta E_1+\Delta E_2)\Delta t}{\Delta E_1\Delta E_2
  (\Delta E_1-\Delta E_2)}.
\end{equation}
For $0\le E\le 0.70$ one obtains a relative deviation of $1.5\cdot 10^{-8}$,
for $0\le E\le 0.92$ the maximal deviation is $5\cdot 10^{-7}$. Finally, the
adjustment of
\begin{equation}
t=t_0+A(E-\erf(t_0))+B(E-\erf(t_0))^2+C(E-\erf(t_0))^3
\end{equation}
leads to
\begin{eqnarray}\label{ABC}
A&=&(3\Delta E_1^2\Delta E_2^2(\Delta E_1-\Delta E_2)
  -2\Delta E_1^2\Delta E_3^2(\Delta E_1-\Delta E_3)\strut\nonumber\\&&\strut
  +\Delta E_2^2\Delta E_3^2(\Delta E_2-\Delta E_3))\Delta t/D,\nonumber\\
B&=&(-3\Delta E_1\Delta E_2(\Delta E_1^2-\Delta E_2^2)
  +2\Delta E_1\Delta E_3(\Delta E_1^2-\Delta E_3^2)\strut\nonumber\\&&\strut
  -\Delta E_2\Delta E_3(\Delta E_2^2-\Delta E_3^2))\Delta t/D,\nonumber\\
C&=&(3\Delta E_1\Delta E_2(\Delta E_1-\Delta E_2)
  -2\Delta E_1\Delta E_3(\Delta E_1-\Delta E_3)\strut\nonumber\\&&\strut
  +\Delta E_2\Delta E_3(\Delta E_2-\Delta E_3))\Delta t/D,
\end{eqnarray}
where $D=\Delta E_1\Delta E_2\Delta E_3(\Delta E_1-\Delta E_2)
(\Delta E_1-\Delta E_3)(\Delta E_2-\Delta E_3)$. For $0\le E\le 0.70$ the
relative deviation is restricted to $5\cdot 10^{-10}$ while up to $E=0.92$
the maximal relative deviation is $4\cdot 10^{-8}$. The results for the
deviations of $T_{(n)}$ ($n=1,2,3$) for linear, quadratic and cubic dynamical
approximation are shown in Fig.~\ref{rval456}.

\section{Discussion}
In order to test the feasibility and speed, we have coded our algorithm in the
computer language C under {\tt Slackware~15.0} ({\tt linux 5.15.19}) on an
ordinary hp laptop with Intel\textregistered\ Core\texttrademark 2 Duo CPU
P8600 @ 2.4GHz with 3MiB memory used. The dependence of the CPU time for the
calculation is estimated by calculating the value $10^6$ times in sequence.
The speed of the calculation of course does not depend on the value for $E$,
as the precision is not optimised. This would have to be neccessary for a
practical application. For an arbitrary starting value $E=0.8$ we perform this
test, and the results are given in Table~\ref{tab1}. An analysis of this table
shows that a further step in the degree $p$ doubles the run time while the
dynamics for increasing $n$ adds a constant value of approximately $0.06$
seconds to the result. Despite the fact that the increase of the dynamics
needs the solution of a linear system of equations and the coding of the
result, this endeavour is justified, as by using the dynamics one can increase
the precision of the result without loosing calculational speed.

\begin{table}[ht]\begin{center}
\caption{\label{tab1}Run time experiment for our algorithm under C for
$E=0.8$ and different values of $n$ and $p$ (CPU time in seconds). As
indicated, the errors are in the last displayed digit, i.e., $\pm 0.01$
seconds.}
\begin{tabular}{|r||c|c|c|c|c|c|}\hline
&$n=0$&$n=1$&$n=2$&$n=3$&$n=4$&$n=5$\\\hline\hline
$p=0$&$0.07(1)$&$0.13(1)$&$0.17(1)$&$0.21(1)$&$0.31(1)$&$0.56(1)$\\\hline
$p=1$&$0.14(1)$&$0.20(1)$&$0.24(1)$&$0.29(1)$&$0.39(1)$&$0.63(1)$\\\hline
$p=2$&$0.25(1)$&$0.32(1)$&$0.35(1)$&$0.40(1)$&$0.50(1)$&$0.75(1)$\\\hline
\end{tabular}\end{center}
\end{table}

The results for the deviations in Figs.~\ref{rval012} and~\ref{rval456} are
multiplied by increasing decimal powers in order to make the result comparable.
This fact indicates that the convergence is improved in each of the steps for
$p$ or $n$ at least by the corresponding inverse power. While the
approximations in the statical approximations $n=0$ in Fig.~\ref{rval012} show
both deviations close to $E=0$ and for higher values of $E$, the dynamical
approximations in Fig.~\ref{rval456} show no deviation at $E=0$ and moderate
deviations for higher values. On the other hand, the costs for an improvement
step in either $p$ or $n$ is at most a factor of two higher CPU time. This
means that the calculation and coding of expressions like Eqs.~(\ref{ABC}) is
justified by the increase of precision gained. Given the goals for the
precision, the user can decide to which degrees $p$ and $n$ the algorithm
should be developed. In order to prove the precision, in Table~\ref{tab2} we
show the convergence of our procedure for $p=2$ fixed and increasing values of
$n$. The last column shows the CPU times for $10^6$ runs of the algorithm
proposed in Ref.~\cite{Dominici:2007} with $N$ given in the last column of the
table in Ref.~\cite{Dominici:2007}, as coded in C.

\begin{table}[hb]\begin{center}
\caption{\label{tab2}Results for $p=2$ and increasing values of $n$ for values
of $E$ approaching $E=1$. The last column shows the CPU time for $10^6$ runs
according to the algorithm proposed in Ref.~\cite{Dominici:2007} for the
values of $N$ given in the last column of the table displayed in
Ref.~\cite{Dominici:2007}.}
\strut\kern-28pt\begin{tabular}{|l||c|c|c|c|c|c||r|}\hline
$E=$&$n=0$&$n=1$&$n=2$&$n=3$&$n=4$&$n=5$&\cite{Dominici:2007}\\\hline\hline
$0.7$&$0.732995$&$0.732868$&$0.732869$&$0.732869$&$0.732869$&$0.732869$
  &$0.17$\\\hline
$0.8$&$0.906326$&$0.906193$&$0.906194$&$0.906194$&$0.906194$&$0.906194$
  &$0.19$\\\hline
$0.9$&$1.163247$&$1.163085$&$1.163087$&$1.163087$&$1.163087$&$1.163087$
  &$0.35$\\\hline
$0.99$&$1.821691$&$1.821376$&$1.821387$&$1.821386$&$1.821386$&$1.821386$
  &$1.95$\\\hline  
$0.999$&$2.326608$&$2.326762$&$2.326752$&$2.326754$&$2.326754$&$2.326754$
  &$14.62$\\\hline
$0.9999$&$2.749217$&$2.751197$&$2.751034$&$2.751076$&$2.751056$&$2.751971$
  &$128.30$\\\hline
\end{tabular}\kern-28pt\end{center}
\end{table}  

\section{Conclusions}
In this paper we developed and described an approximative algorithm for the
determination of the error function which is based on geometric considerations.
Along the lines explained in this paper, the algorithm can be easily
implemented and extended. We have shown that each improvement step gains an
improvement of the precision of at least a factor of ten, at the cost of at
most a factor of two more CPU time. As a bonus, we have given a geometric
derivation of Craig's integral representation of the error function and a
converging power series expansion for this formula.

\subsection*{Acknowledgments}
This research was funded by the European Regional Development Fund under
Grant No.~TK133.

\end{document}